\documentclass[preprint,preprintnumbers,amsmath,amssymb,aps]{revtex4}
\usepackage{graphicx}

\newcommand{\bee}{\begin{equation}}
\newcommand{\ee}{\end{equation}}
\newcommand{\beea}{\begin{eqnarray}}
\newcommand{\eea}{\end{eqnarray}}
\def\Tr{{\rm Tr}}

\begin{document}

\title{Improving meson two-point functions in lattice QCD}

\author{Thomas DeGrand}
\author{Stefan Schaefer}
\affiliation{
Department of Physics, University of Colorado,
        Boulder, CO 80309 USA}

\begin{abstract}
We describe and test a method to
 compute Euclidean meson two-point functions in  lattice QCD.
The contribution from the low-lying eigenmodes of the Dirac operator 
is averaged over all positions of the quark sources. The contribution
from the higher modes is estimated in the traditional way with one
or a few source points per lattice.
 In some channels, we observe a significant improvement 
in the two-point functions for  small quark masses.
\end{abstract}
\maketitle

\section{Introduction}
In current lattice QCD computations a significant amount of effort is put
into the simulation at small quark masses. Unfortunately, the noise
in the meson correlators increases with decreasing quark mass. The purpose
of this paper is to examine a method to reduce this noise. We are
motivated by observations 
that for lighter quarks the low-lying eigenmodes of the
Dirac operator make an important contribution to hadron correlators
in some channels.
Traditionally one restricts oneself to the computation of propagators using
 a single position of the quark source.
However, by computing the low-lying eigenmodes, one can average 
their contribution over all possible positions of the source and hope 
that this improves the signal. The contribution from the high modes can
be estimated in the  standard way.

\begin{figure}
\begin{center}
\includegraphics[width=0.3\textwidth,clip,angle=-90]{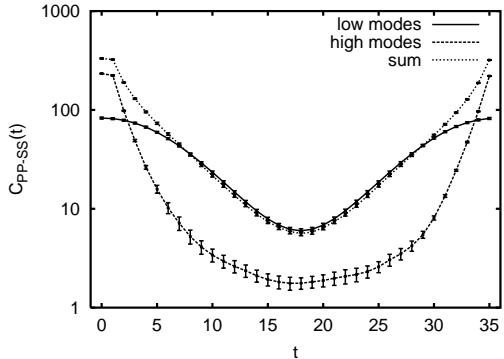}
\end{center}
\caption{\label{fig:0} Pseudoscalar-scalar difference correlator
at $am_q=0.020$,
from the $a=0.13$ fm $12^3\times 36$ data set used in this study,
 showing the contribution of the lowest 20 eigenmodes of $D^\dagger D$,
the contribution of the high modes, and the full correlator.}
\end{figure}

The saturation of the 
correlators has, for example, been studied in \cite{DeGrand:2000gq,Neff:2001zr}.
 Refs \cite{DeGrand:2000gq,DeGrand:2003sf} have shown that for
 actions similar to the one
used in this study, the long distance part of many meson correlators is dominated
by the low eigenmode contribution to the quark propagator. Fig. \ref{fig:0}
shows an example of this behavior. This is the pseudoscalar-scalar
difference correlator evaluated at a quark
mass which is about $0.3m_{strange}$, with a Gaussian source and a point sink.

We wish to calculate a correlator
\bee
C(t) = \frac{1}{L^3T}\sum_{x',x'',t''}\langle O_1(x',t'' + t) O_2(x'',t'') \rangle .
\label{eq:C}
\ee
Inserting a complete set of relativistically normalized states this becomes up
to boundary effects,
\bee
C(t) = \sum_n \frac{\langle 0 | O_1 |n\rangle\langle n | O_2|0\rangle}{2m_n}e^{-m_nt},
\ee
from which a fit allows us to read off masses and matrix elements.
For bilinear operators, $O_i = \bar \psi \Gamma_i \psi$,
\bee
C(t)= \frac{1}{L^3T}\sum_{x',x'',t''} \langle \Tr \Gamma_1 G(x',t''+t;x'',t'')\Gamma_2
     G(x'',t'';x',t')\rangle
\label{eq:GG}
\ee
where $G(x',t';x,t)$ is the quark propagator, $(D+m)G(x',t';x,t)=\delta^3_{x,x'}\delta_{t,t'}$.
Evaluating $G$ in a way that it can be usefully combined in Eq.~\ref{eq:GG} requires
a separate inversion of $(D+m)$ for every source point $(x,t)$.
Usually, lattice calculations replace the $1/(L^3T)\sum_{x'',t''}$ of Eq.~\ref{eq:C}
by one or a few points per lattice,
\bee
C_1(t)=\sum_{x',x'',t''}\langle O_1(x',t'' + t) O_2(x'',t'') \rangle.
\label{eq:C1}
\ee
Obviously, $C_1(t)$ is an approximation to $C(t)$ on a single background configuration, so
in an average over an ensemble of gauge configurations
 $\langle C_1(t) \rangle = \langle C(t) \rangle+O(1/\sqrt{L^3T})$.

If preparation of lattices for use in a data set were expensive (for example, in a dynamical fermion
simulation) it would be a more efficient use of them to compute correlators like
$C(t)$ rather than $C_1(t)$. If it were possible to do part of the calculation
cheaply, it might also be advantageous to break up the calculation
\bee
C(t)= C_A(t)+C_B(t)+\dots
\ee
and approximate only some parts of the correlator (only $C_A(t)$, say) by a $C_1$.
This can be done easily if one has constructed $n$ eigenmodes of the massless Dirac
operator $D|j\rangle = i\lambda_j |j\rangle$. Then the quark propagator
can be broken into two pieces, $G=G_L +G_H$,
where $G_L$ is given by an explicit mode sum
\bee
G_L(x,t;x',t')= \sum_{j=1}^n \frac{\langle x,t|j\rangle\langle j|x't'\rangle}{i\lambda_j+m}.
\ee
and $G_H$ is the remainder.
Then the meson correlator becomes
\bee
C(t)= C_{LL}(t)+  C_{HL}(t) + C_{LH}(t)+ C_{HH}(t).
\label{eq:breakup}
\ee
$C_{LL}$ would be calculated from all source points on the lattice, while
$C_{LH}$, $C_{HL}$, and $C_{HH}$ would come from only a few source points, or a single point.
Note that as long as one does not include all eigenmodes into the sum, either directly
or indirectly through Eq.~\ref{eq:breakup}, one is only approximating the two point function. 
$C(t)$ does not equal $ C_{LL}(t)$ alone.

Does this trick gain anything in statistics? One might suspect that it would,
for several reasons. First, there is just more sampling per lattice. If we
could do a full ``all-to-all'' calculation (with no mode truncation) we might expect some
gain due to the larger data sample. It would not be a complete reduction by $1/\sqrt{L^3T}$,
because correlators from nearby sources on the lattice are highly correlated,
but  some gain might result.
Second, it is reasonably well-known that at low quark masses, hadron propagators from a single
source
do not show a regular exponentially-falling behavior.
Some averaging over position is needed to smooth the signal. This can be done by
using several propagators for several source points, at a cost which increases linearly
per source point. However, most of the irregularity comes from the low eigenmode
part of the propagator. Computing ``all to all''propagators from the low mode part
of the quark propagator is essentially free once the low modes are computed.

Using Eq.~\ref{eq:breakup} at larger quark mass is not likely to improve the signal,
simply because low modes do not make a significant contribution to any meson correlator.
At larger quark mass, $C_{LL}(t)$ remains very ``flat'' while $C(t)$ falls steeply.
There is a considerable cancellation between $C_{LL}(t)$, $C_{HH}(t)$, and the interference terms.
The noise in this cancellation is not reduced by averaging $C_{LL}(t)$ alone.

The method has an obvious additional cost, compared to a usual spectroscopy simulation:
the eigenmodes must be computed. This may or may not be a significant overhead.
For the data set of our tests,  computing the lowest twenty eigenmodes of the squared massless
Dirac operator takes about 8 time units, while the complete set of
 quark propagators
from the lightest mass  studied to the heaviest takes about 16 time
 units times two (for two sets
of propagators) per lattice. These eigenmodes were used to precondition the inversion of
the Dirac operator, and this appeared to reduce the number of Conjugate Gradient steps needed 
at the lowest quark mass by about forty per cent. Even doing one source point of
propagators is
accelerated by projecting out low modes.

We are not aware of previous applications of this rather simple idea.
It is common to use low eigenmodes of the Dirac operator directly in overlap fermion calculations.
(Compare the recent work by Giusti, Hernandez, Laine, Weisz, and Wittig\cite{Giusti:2003iq}.)
DeGrand and Heller\cite{DeGrand:2002gm} used a similar ``hybrid''  method to compute disconnected diagrams,
summing the low mode contribution exactly and using a stochastic estimator for
the high mode part of the correlator.
 ``All-to-all''
quark propagators have been computed using rather different (stochastic)
 methods by Duncan, Eichten and Yoo \cite{DEY}.
We believe that C.~Michael and S.~Wright\cite{WRIGHT} are doing similar work.

\section{Numerical tests}
In order to test these statements we used quark propagators and eigenmodes used in 
a previous measurement of the $B_K$ parameter \cite{DeGrand:2003in}.
They were computed on 80 quenched 
gauge configurations of size $36\times 12^3$ generated with the 
Wilson gauge action at $\beta=5.9$. The fermion action is an 
overlap \cite{ref:neuberfer} Dirac operator with a kernel action which uses first and second
nearest neighbor interactions \cite{ref:TOM_OVER}
 which are themselves composed of  HYP blocked \cite{ref:HYP}
gauge links. The lowest twenty eigenmodes of the massless
 overlap Dirac operator $D(0)$ are constructed from eigenmodes
of the squared Hermitian Dirac operator $H(0)^2=D(0)^\dagger D(0)$ with  $H(0)=\gamma_5 D(0)$,
 using an adaptation of
a Conjugate Gradient algorithm of Bunk et al.
and  Kalkreuter and Simma \cite{ref:eigen}. We consider spectroscopy with
four quark masses $m_q=0.015$, $0.020$, $0.025$ and $0.035$,
which correspond to pseudoscalar to vector meson mass ratios 
 $m_{\rm PS}/m_{\rm V}$ ranging between about 0.4 to 0.64.
The correlators used Gaussian sources  $\Phi=\exp(-r/r_0)^2$ with a size $r_0/a=3$.
All the data shown will use point sinks projected onto zero momentum by
summing over each time slice.
 On each of the configurations the inversion 
of the Dirac operator was done on two sources, one located on time-slice
$t=0$, the other on $t=16$. We average over these two positions.
The $\Gamma$ matrices depend on the meson in question and we shall use the
following abbreviations:
\begin{table}[h]
\begin{center}
\begin{tabular}{c|c|c|c|c|c}
P& S& $A_\mu$&$ V_\mu$ &$B_{\mu\nu}$\\
\hline
$\gamma_5$ &$ 1$ &$ \gamma_5\gamma_\mu$ &$\gamma_\mu$ &$\gamma_\mu\gamma_\nu$
\end{tabular}
\end{center}
\end{table}

In order to remove the zero modes from the pion correlator we  consider 
 primarily the difference of the pseudo-scalar and the scalar 
correlator ($PP-SS$).

\begin{figure}
\begin{center}
\includegraphics[width=0.3\textwidth,clip,angle=-90]{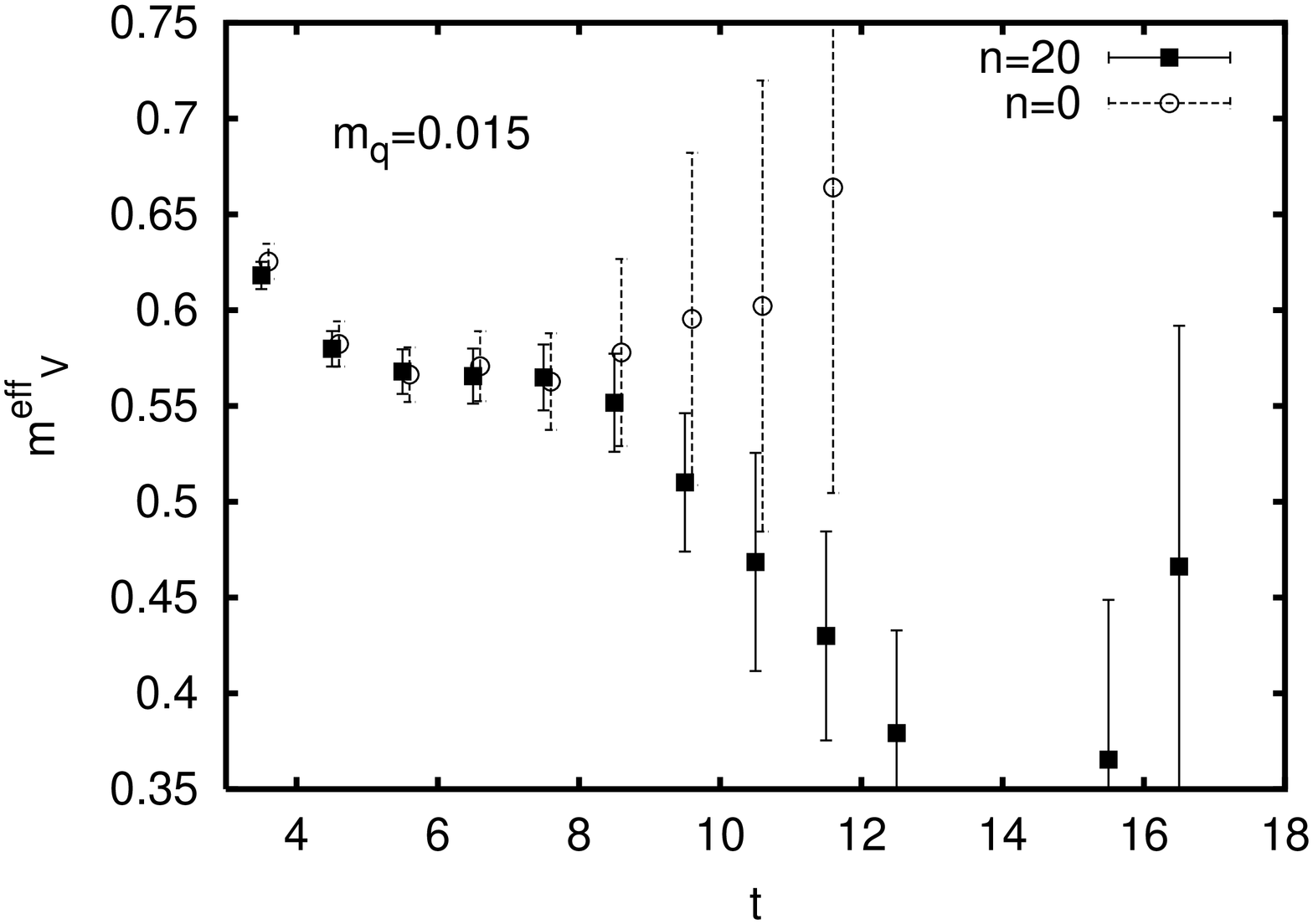}
\includegraphics[width=0.3\textwidth,clip,angle=-90]{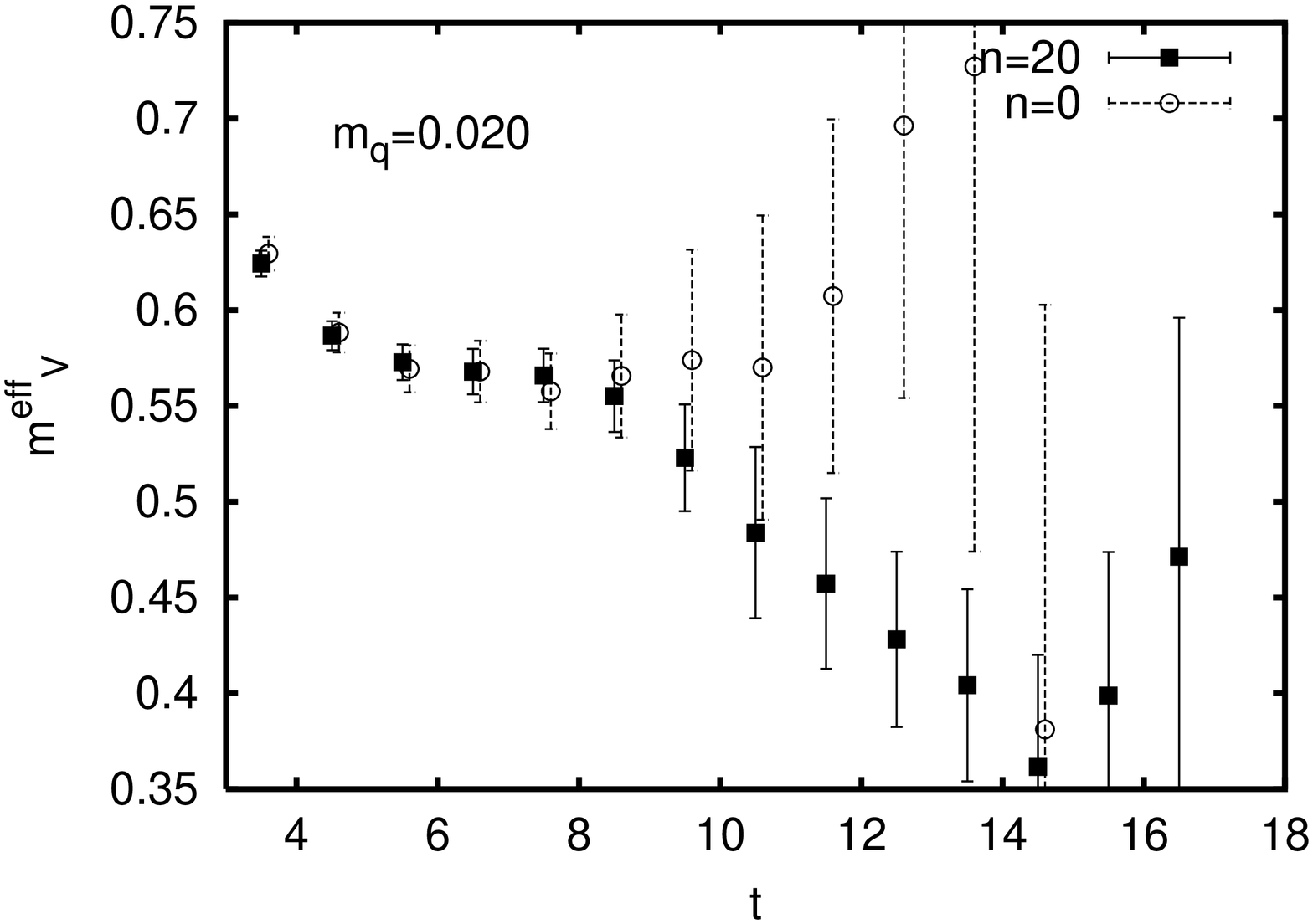}
\includegraphics[width=0.3\textwidth,clip,angle=-90]{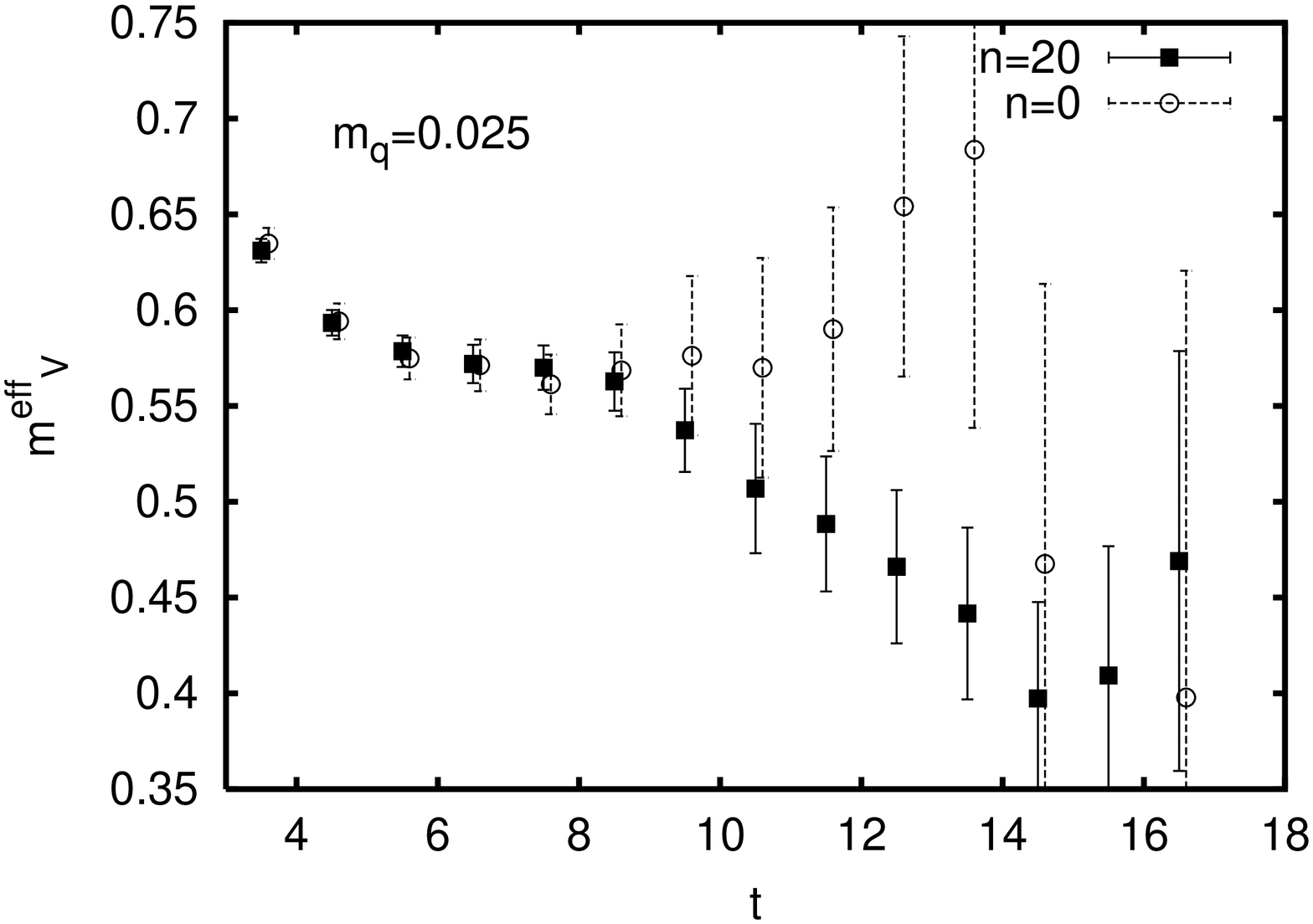}
\includegraphics[width=0.3\textwidth,clip,angle=-90]{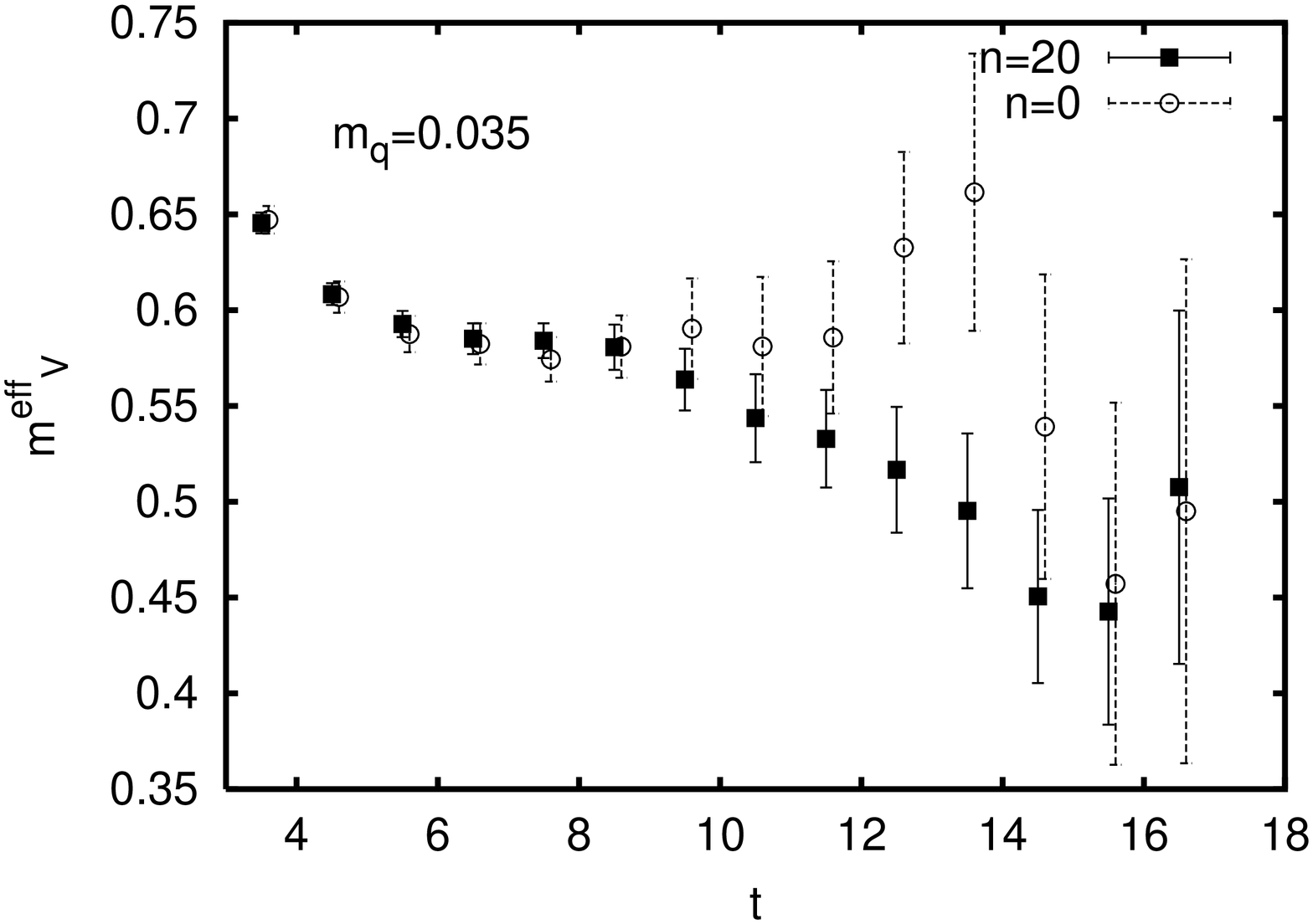}
\end{center}
\caption{\label{fig:3}Effective mass of the $VV$ 
correlator for different quark masses.}
\end{figure}

To begin our study of the effect of keeping all source points for the low
eigenmodes,  we show the ratio of the 
error bars for $C(t)$
at fixed time-slice $t=5$ for $n=20$ eigenmodes included as ``all-to-all''
propagators and $n=0$ in Fig.~\ref{fig:3}.
We observe a gain of up to $30\%$. 
The gain is also larger if one does not average over the two 
sources for each configuration. 
For correlators which are well behaved for $t$ closer to $T/2$, the
gain in this region is also larger as the low-lying modes become
more important there.

\begin{figure}
\begin{center}
\includegraphics[width=0.3\textwidth,clip,angle=-90]{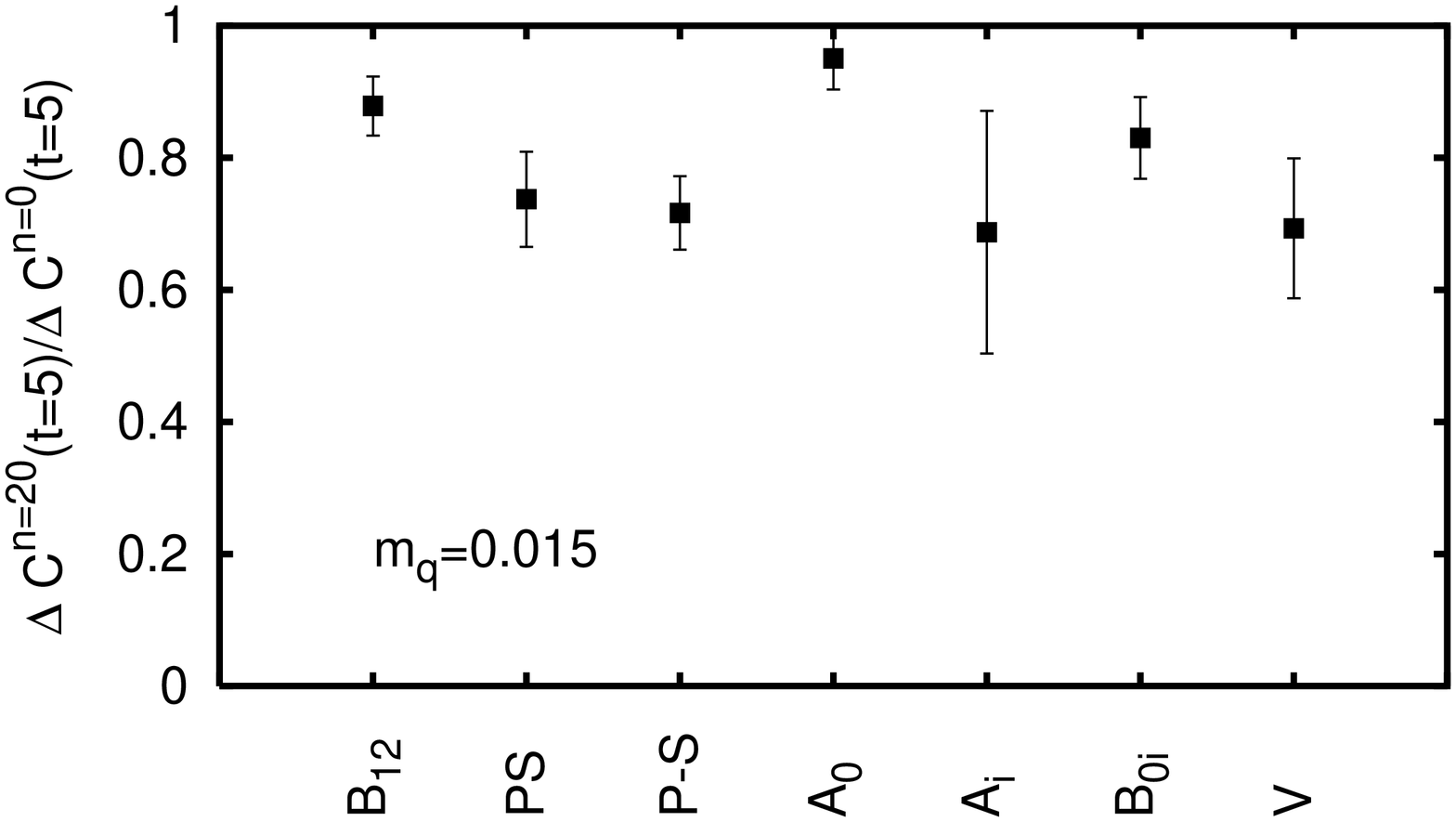}
\includegraphics[width=0.3\textwidth,clip,angle=-90]{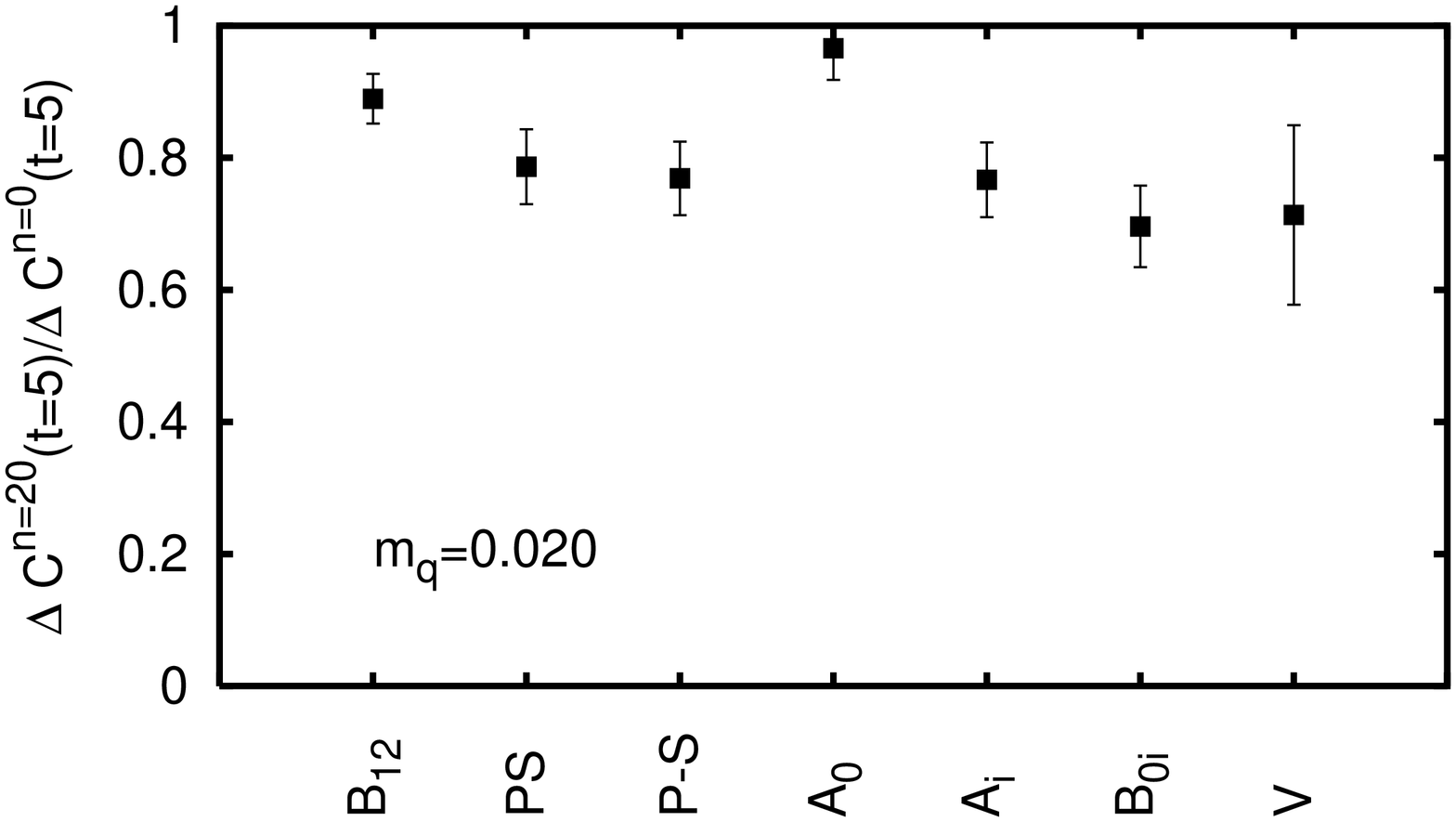}
\includegraphics[width=0.3\textwidth,clip,angle=-90]{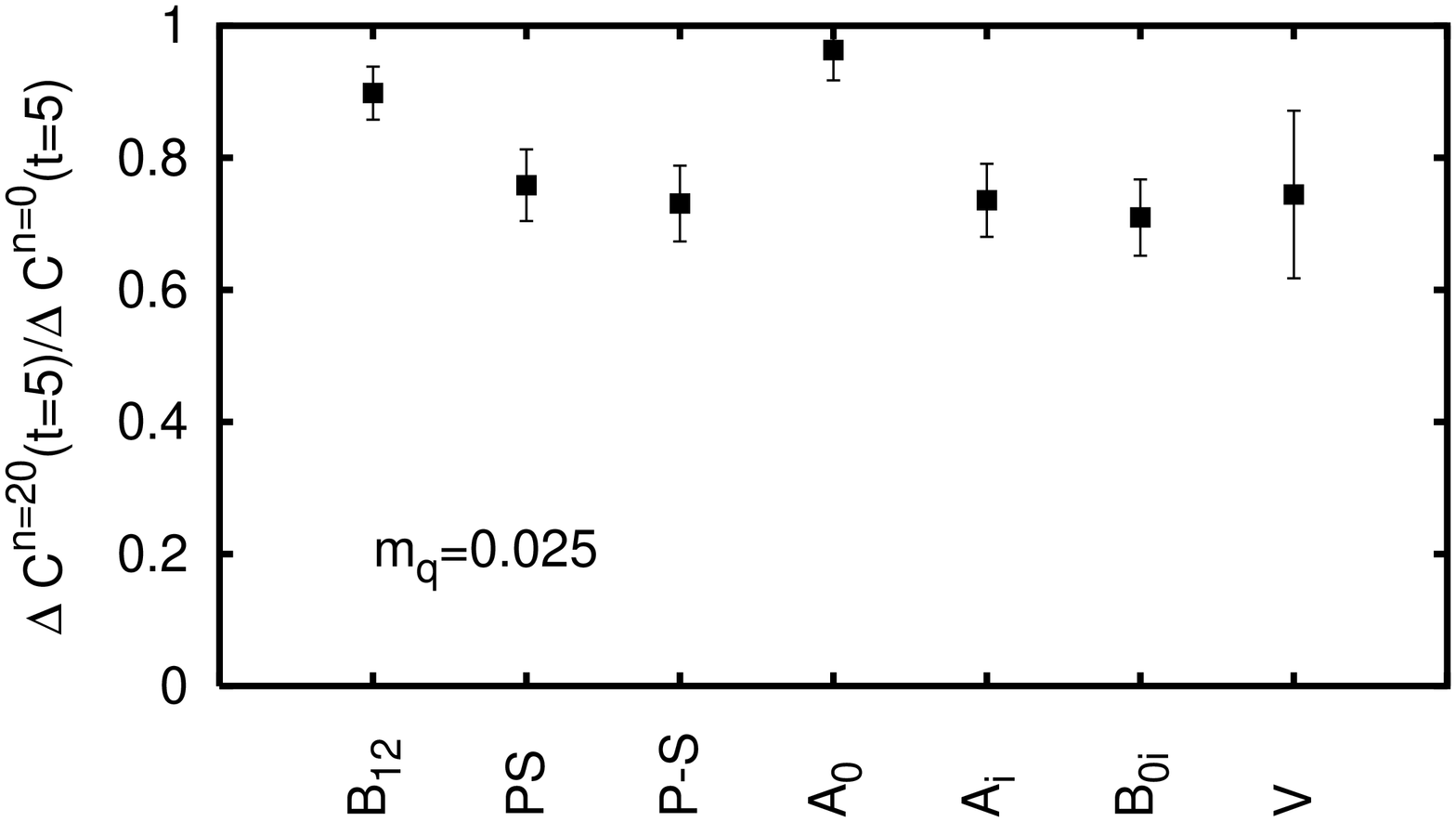}
\includegraphics[width=0.3\textwidth,clip,angle=-90]{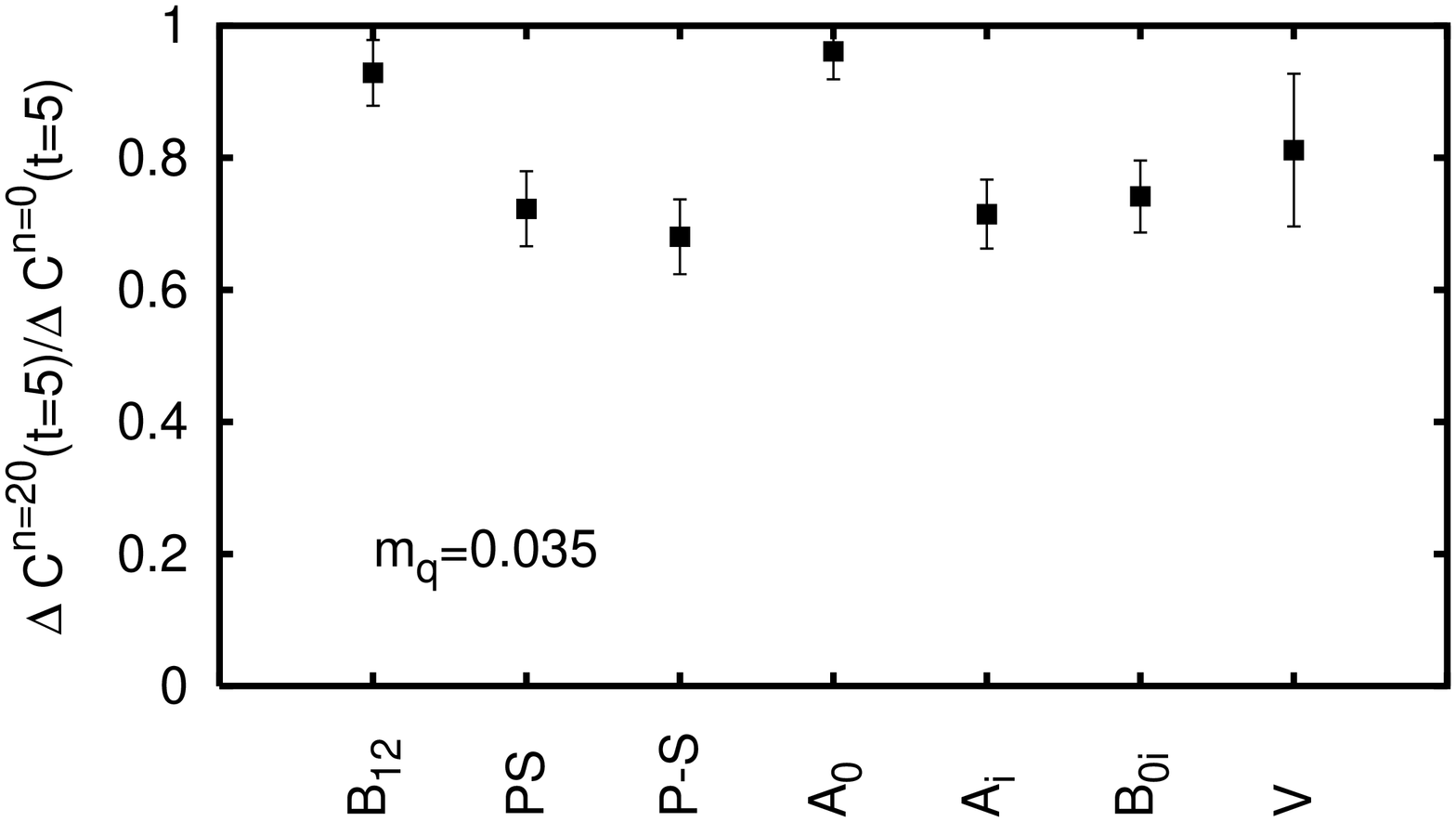}
\end{center}
\caption{\label{fig:5} Ratio of the error bars for $n=20$ compared $n=0$ at 
time-slice $t=5$ for the different meson correlators. 
Data uses two source points for the high eigenmode part of the correlator.
}
\end{figure}

\begin{figure}
\begin{center}
\includegraphics[width=0.3\textwidth,clip,angle=-90]{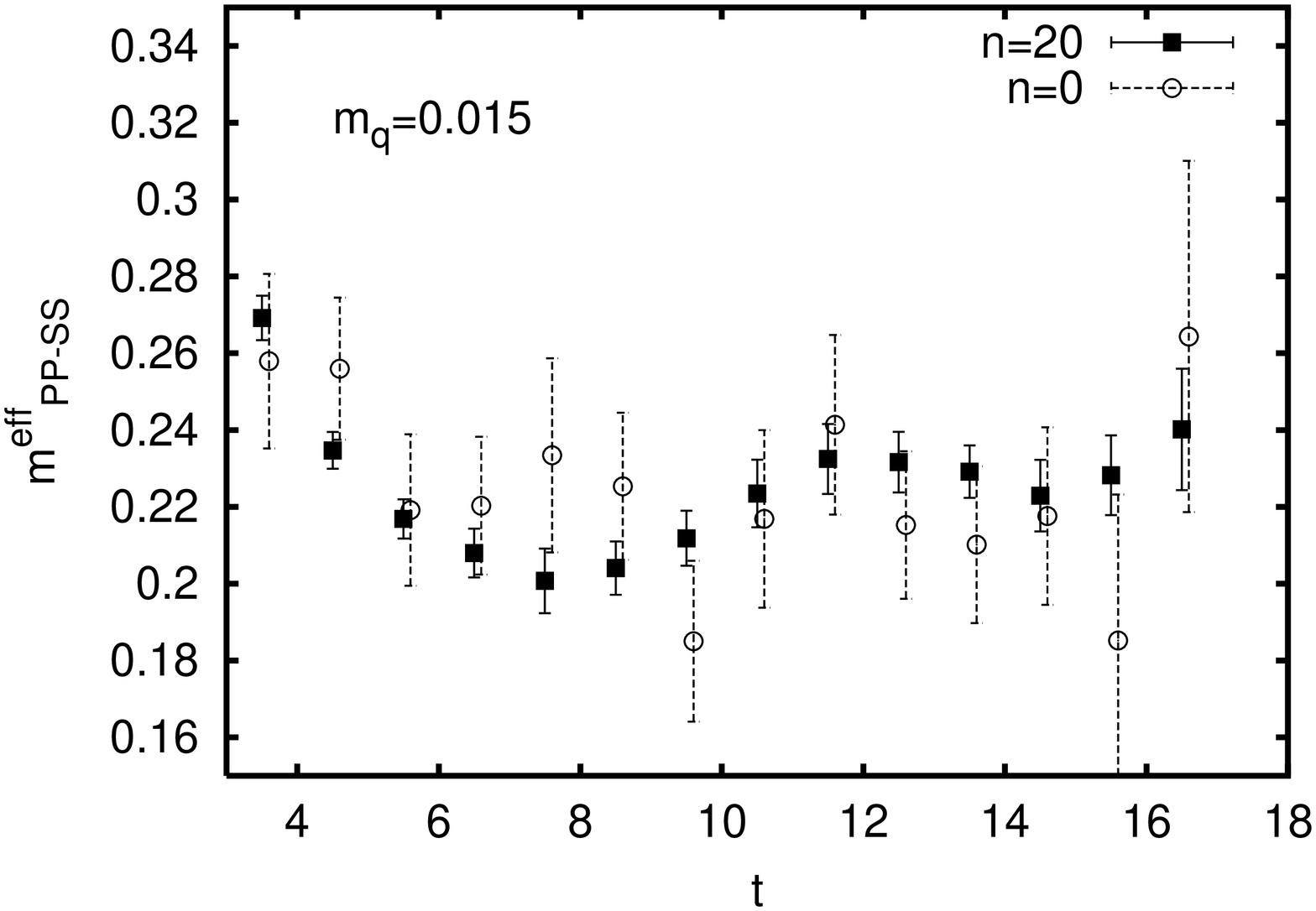}
\includegraphics[width=0.3\textwidth,clip,angle=-90]{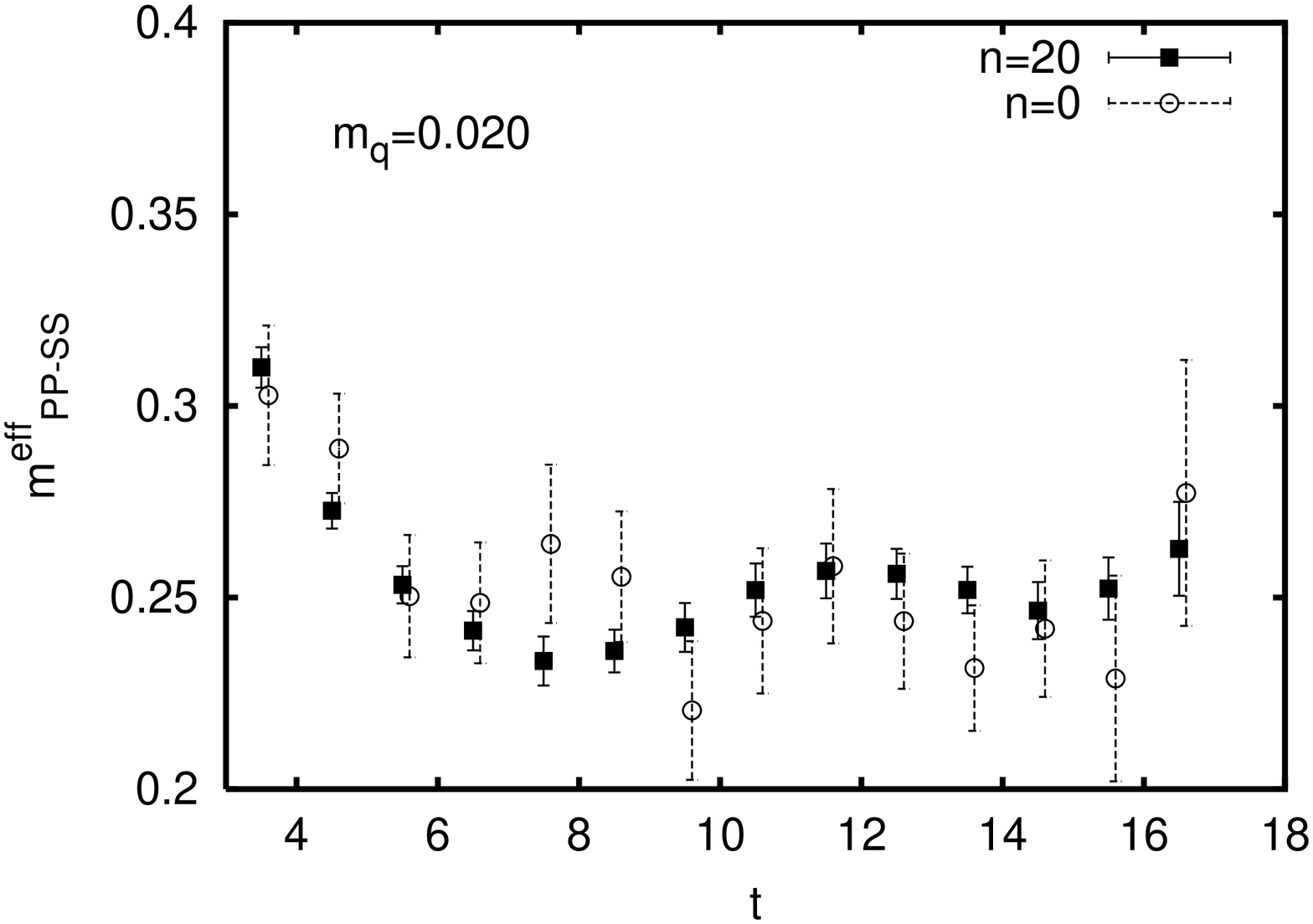}
\includegraphics[width=0.3\textwidth,clip,angle=-90]{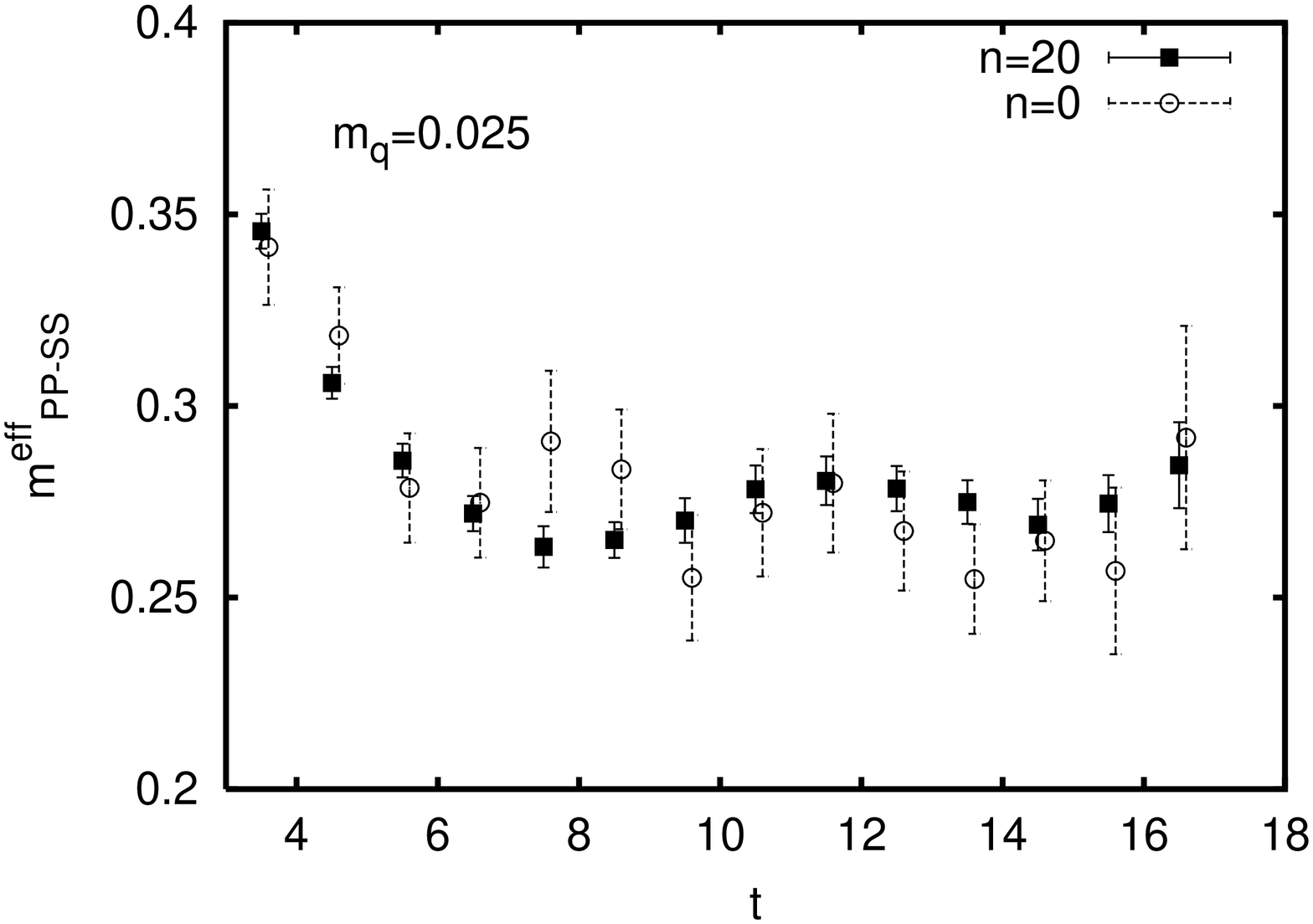}
\includegraphics[width=0.3\textwidth,clip,angle=-90]{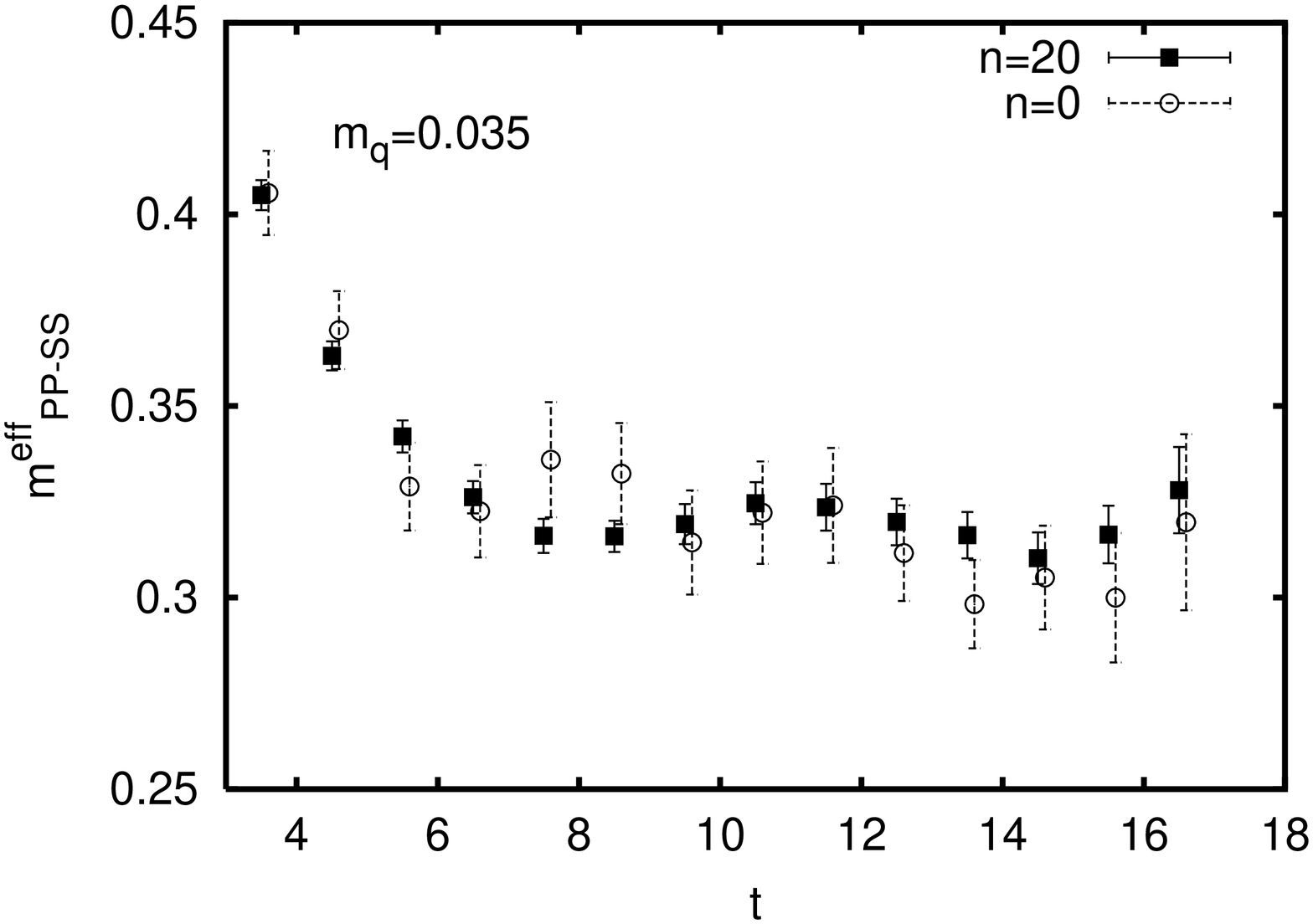}
\end{center}
\caption{\label{fig:1}Effective mass of the $PP-SS$ 
correlator for different quark masses.}
\end{figure}

Next, we turn to spectroscopy.
In Fig.~\ref{fig:1} we show the effective mass plots for the $PP-SS$ correlator
and the vector meson. $n=0$ is the correlator computed with the conventional method,
whereas $n=20$ labels the one which the average over the sources taken over the 20 lowest
eigenmodes.  We observe a significant improvement in the
error bars of the two-point functions.
The jackknife error bars for both methods are consistent.

\begin{figure}
\begin{center}
\includegraphics[width=0.3\textwidth,clip,angle=-90]{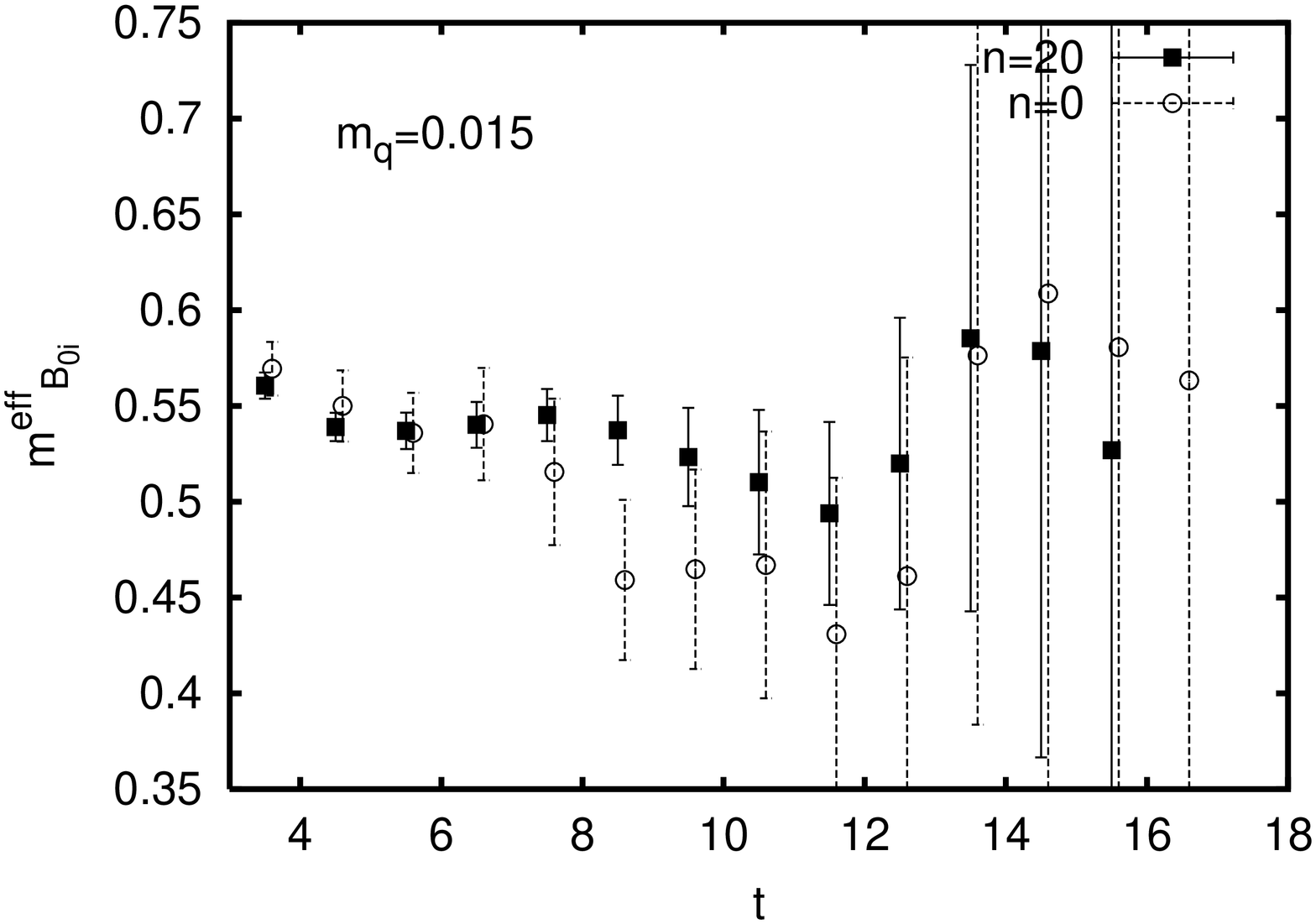}
\includegraphics[width=0.3\textwidth,clip,angle=-90]{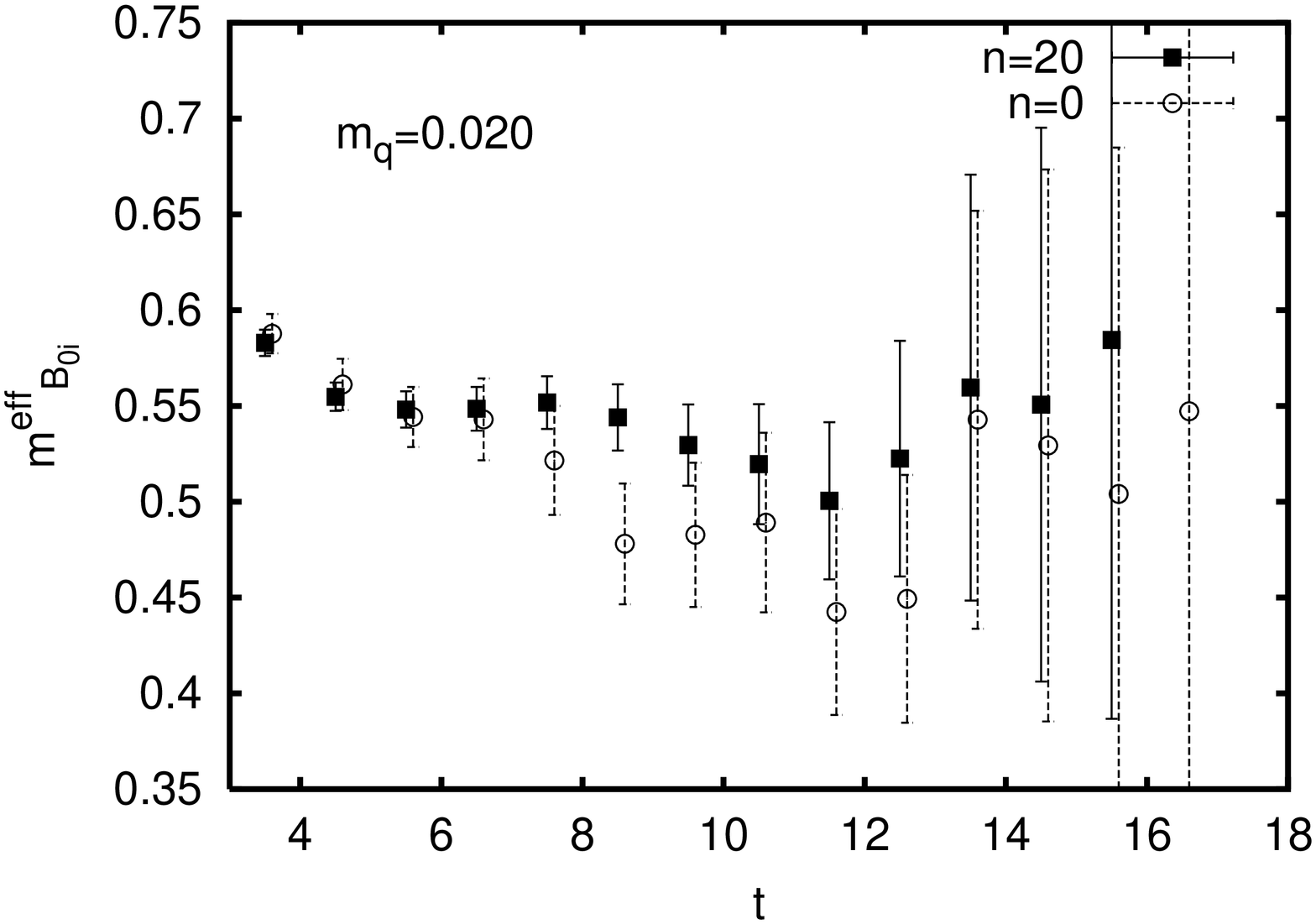}
\includegraphics[width=0.3\textwidth,clip,angle=-90]{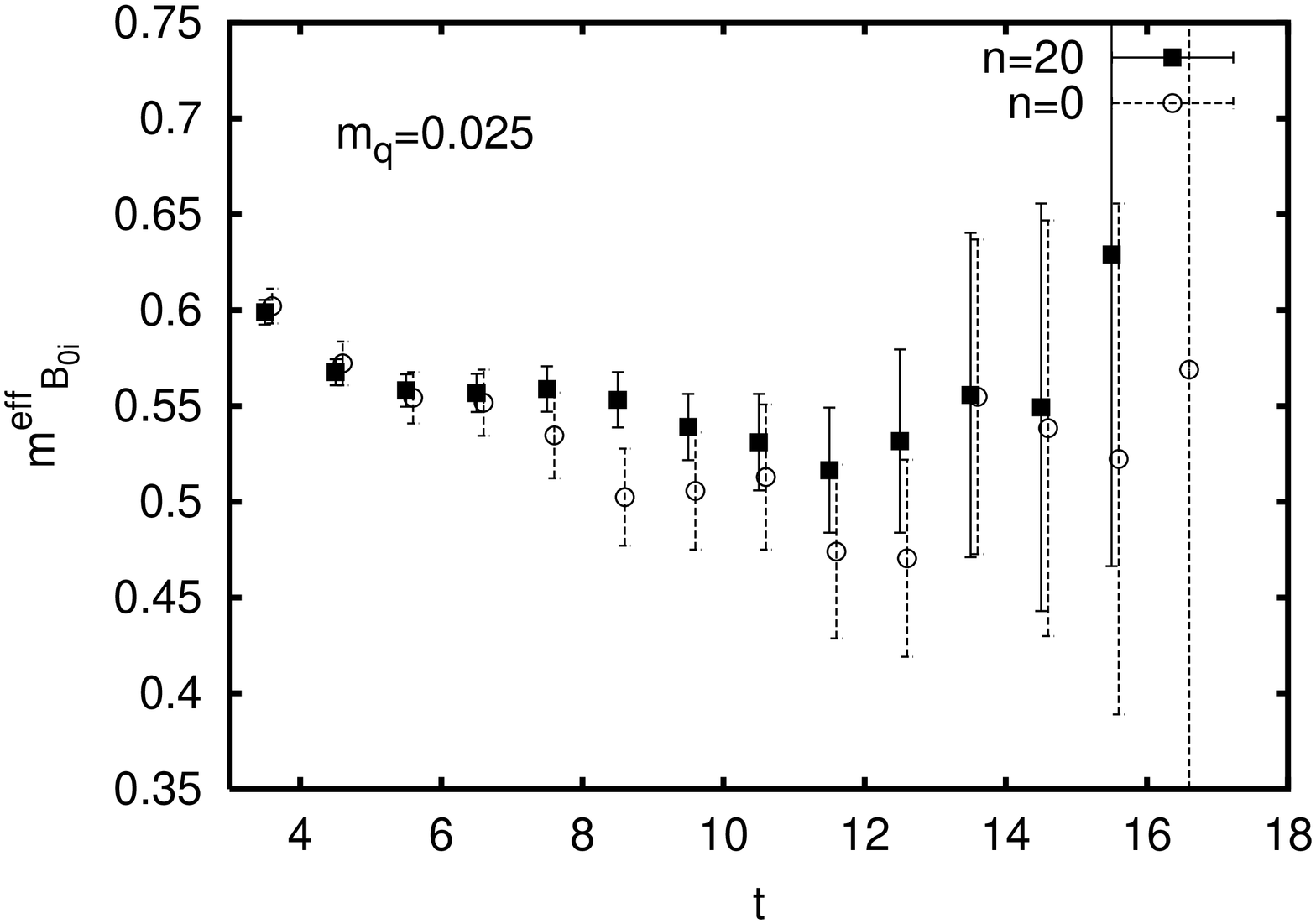}
\includegraphics[width=0.3\textwidth,clip,angle=-90]{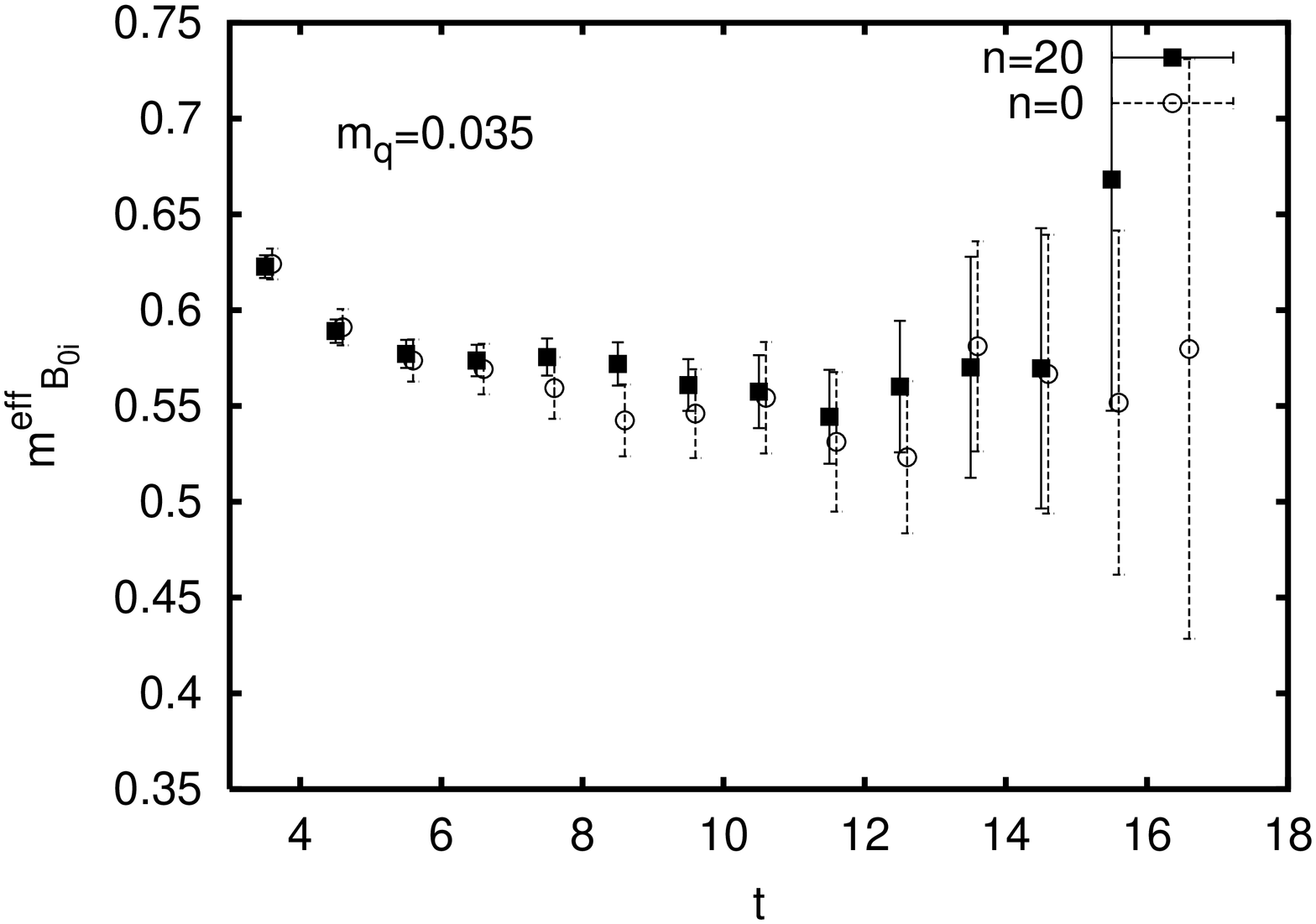}
\end{center}
\caption{\label{fig:2}Effective mass of the $B_{0i}$ 
(vector) correlator for different quark masses.}
\end{figure}

In Figs.~\ref{fig:2} and \ref{fig:3} we show the effective mass plots
for the vector meson extracted from the $VV$ and the $B_{0i}B_{0i}$
correlators. In both channels large fluctuations at large $t$ due to the small
box size are visible. The improvement in the two-point function
due to the eigenmodes is small in the $VV$ channel. In the
$B_{0i}B_{0i}$ however the signal improves significantly. (This
channel has a contribution  when both the quark and antiquark propagate through
zero modes, as opposed to the $VV$ channel, which has only zero mode-nonzero mode
interference terms.) The masses extracted from the two
correlators are consistent.

\begin{figure}
\begin{center}
\includegraphics[width=0.3\textwidth,clip,angle=-90]{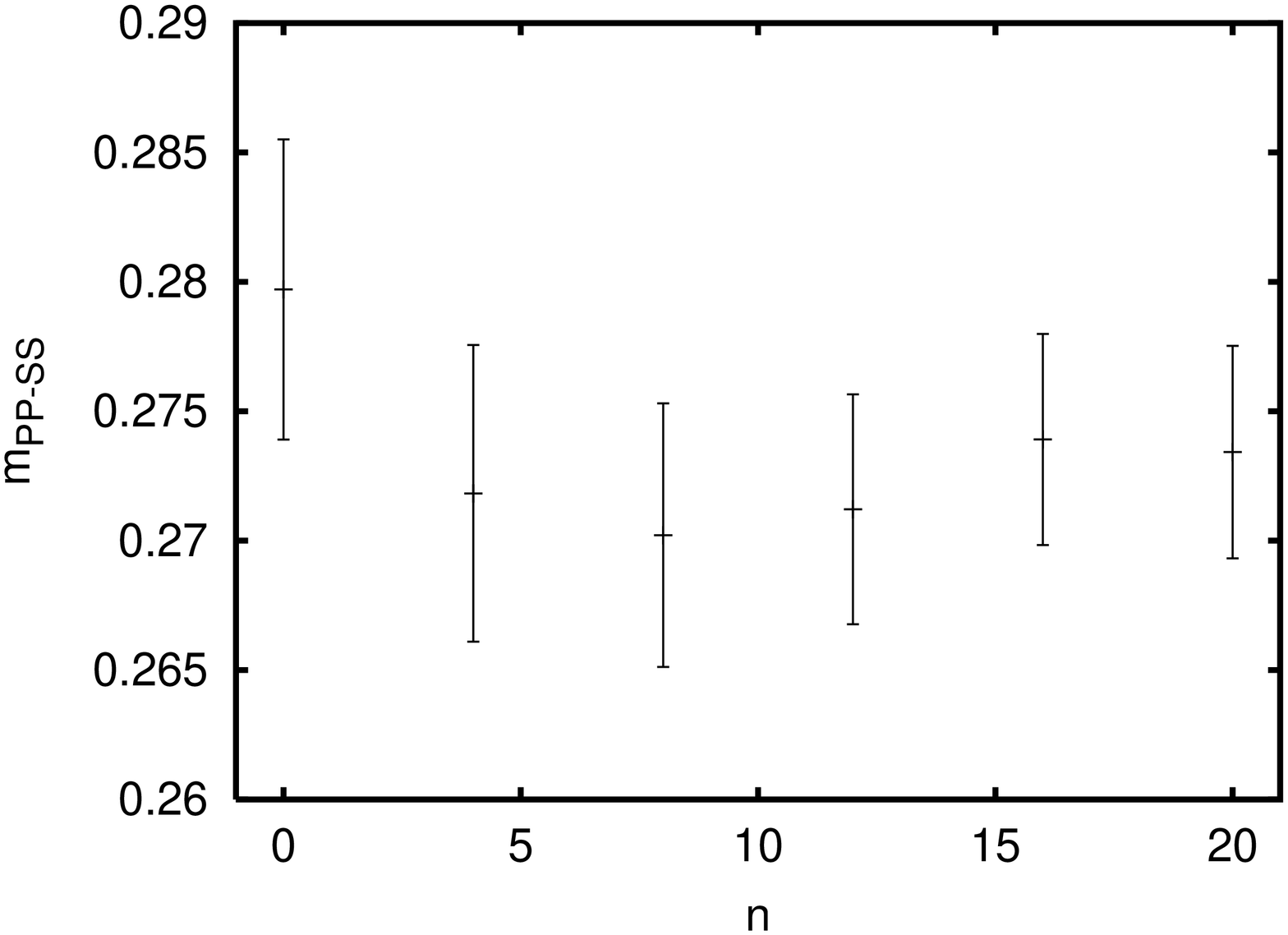}
\includegraphics[width=0.3\textwidth,clip,angle=-90]{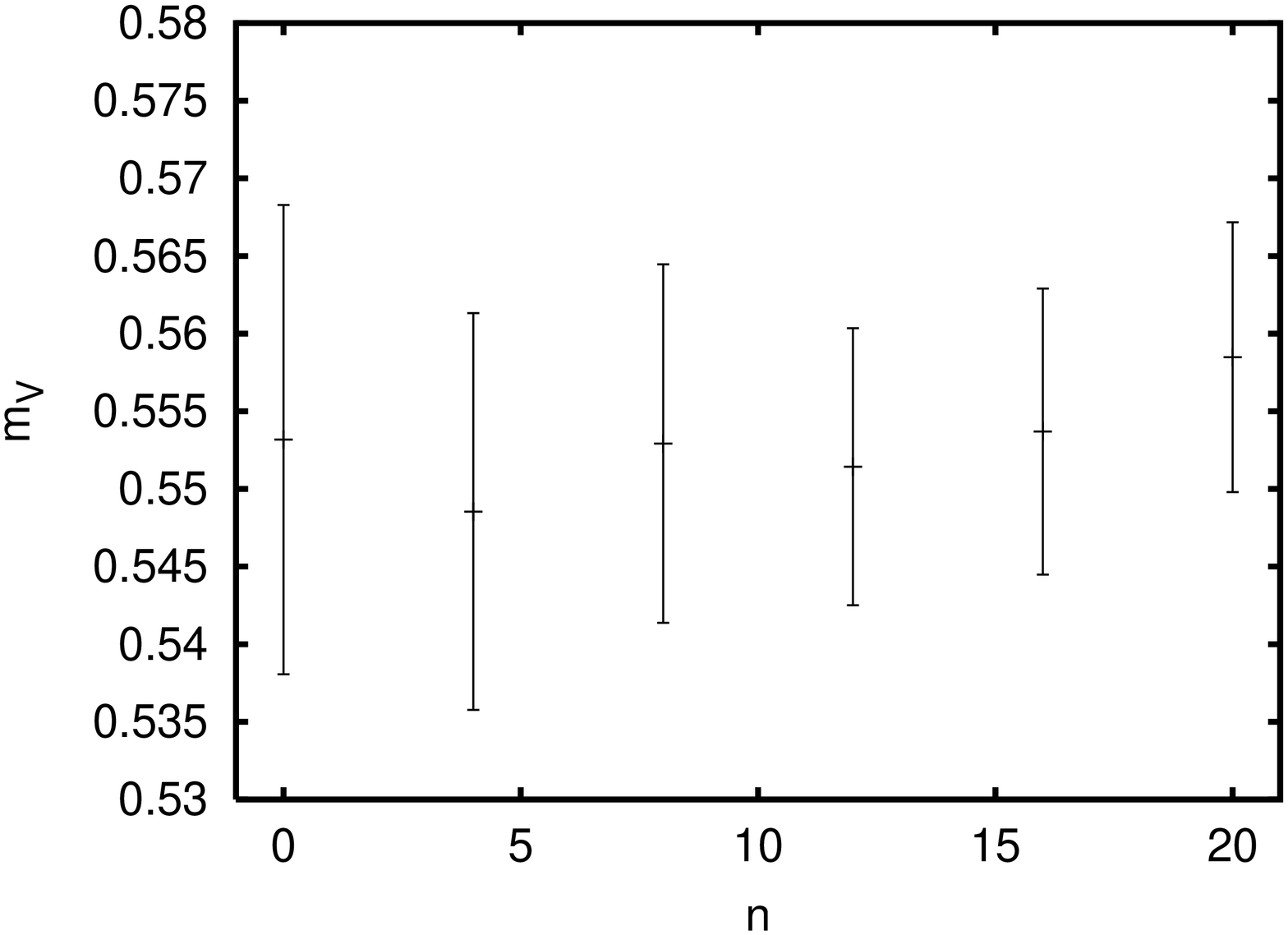}
\end{center}
\caption{\label{fig:6} Dependence of the extracted $PP-SS$ and $B_{0i}$ 
mass on the number of 
eigenmodes included at $a m_q=0.025$. 
}
\end{figure}

The question arises, how many eigenmodes are sufficient to achieve the
desired improvement in the error bar. (Obviously, the answer will depend on quark mass,
simulation volume, and also probably on lattice spacing and choice lattice fermion discretization.)
In Fig.~\ref{fig:6} we show the dependence of the extracted mass from the $PP-SS$
and $B_{0i}$ correlator on the number of eigenmodes included. 
For the $PP-SS$ the gain in the error bar is negligible 
when only $4$ eigenmodes are included.  This
is understandable as we have typically 3 to 4 zero modes per configuration, and these
channels do not couple to zero modes. 
The gain is $23\%$ for 12 and $30\%$ for 16 and 20 modes
included. For the vector meson mass from the $B_{0i}$ correlator, 
the gain is almost linear in from $0$ to 12 included 
eigenmodes and constant at $40\%$ thereafter. In this case
we gain   because the correlator has contributions from
zero modes.

\begin{figure}
\begin{center}
\includegraphics[width=0.3\textwidth,clip,angle=-90]{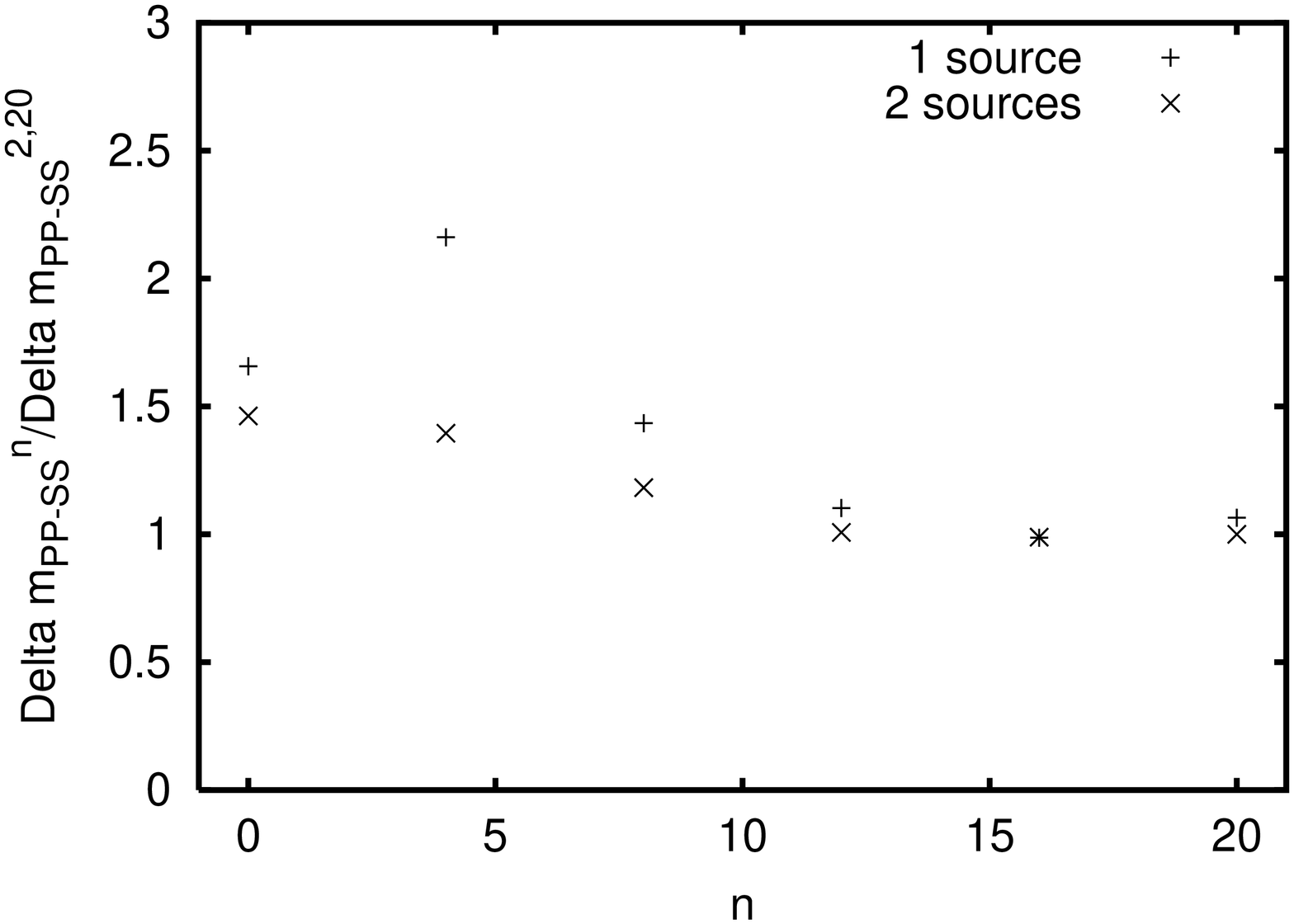}
\includegraphics[width=0.3\textwidth,clip,angle=-90]{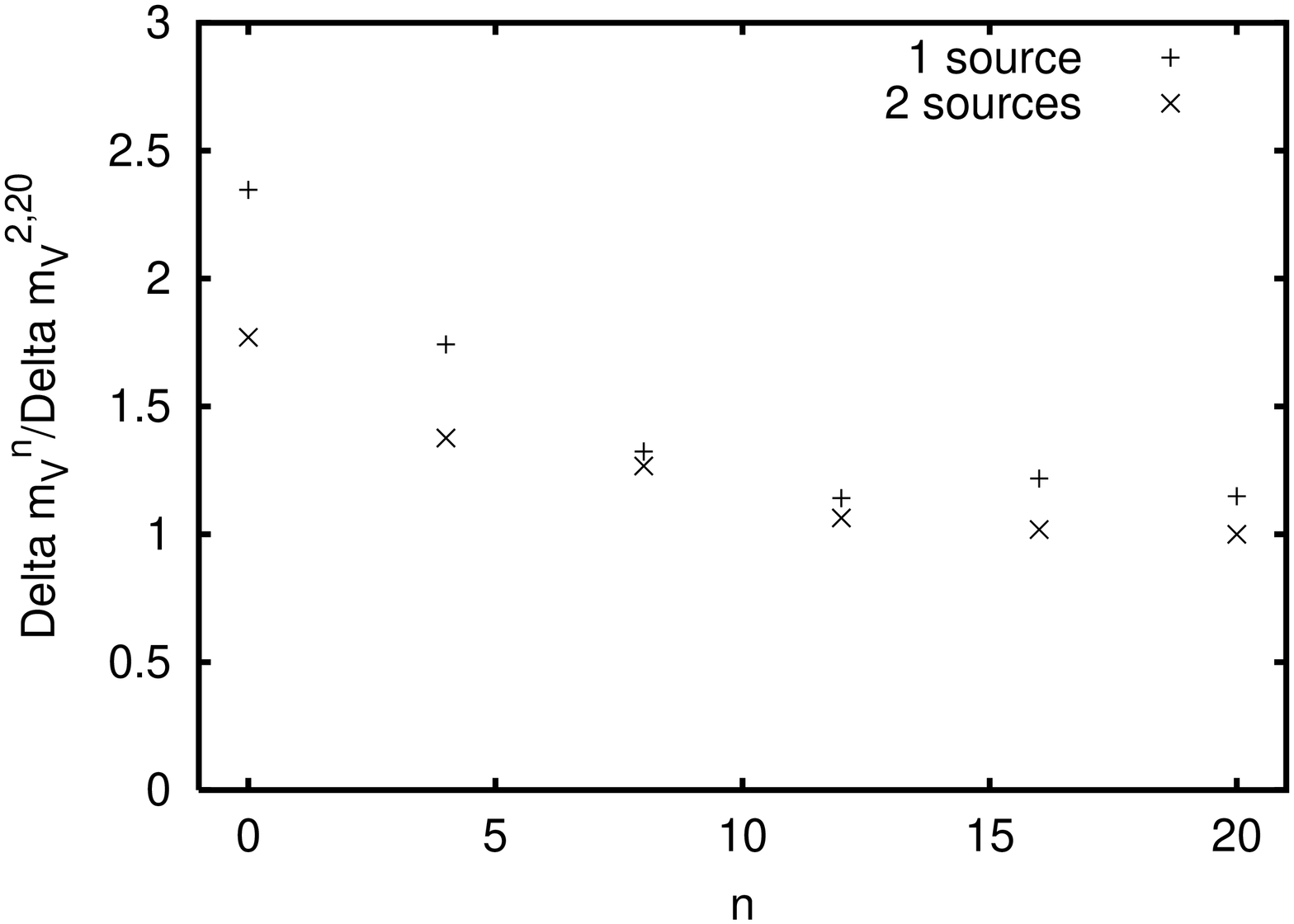}
\end{center}
\caption{\label{fig:7} The error of the  $PP-SS$ and $B_{0i}$ 
mass for one and two source points normalized to the error
for the mass extracted from the correlator with two source points
and $20$ eigenmodes included. Again, $am_q=0.025$ and the fit range
was from $t=8$ to 18.
}
\end{figure}

Let us finally look at the interplay between the inclusion of
a second source  and the eigenmodes.
In Fig.~\ref{fig:7} we plot the uncertainty for the masses divided
by the uncertainty for the mass extracted from the correlator with
two sources and 20 eigenmodes. We see that the uncertainty
of $m_{PP-SS}$ ($m_V$) decreases by about $12\%$ ($30\%$) by
including the second source. However, by including a sufficiently large number of 
eigenmodes, the effect of the second source becomes
smaller, i.e. a  $6\%$ and $14\%$ gain, respecively.
 
\section{Summary}
To summarize, replacing the contribution of low eigenmodes to hadron propagators by
an ``all to all'' contribution  significantly improves the quality of the signal in some channels.
 This is especially the case for the pseudoscalar-scalar difference
and the $B_{0i}$ vector meson correlator.
We demonstrated that for these channels this gain translates into
a smaller uncertainty in the meson masses. 

We studied the dependence of 
their error bars on the number of eigenmodes included and found
that for our simulation parameters it is sufficient to include
about 12 eigenmodes. This number is presumably simulation volume dependent.
If one had a set of eigenmodes in hand, it would be easy to test how
many eigenmodes
would be needed, and over what quark mass range the method would reduce
fluctuations enough to be worth pursuing. This could be done with
 low statistics investigations, basically by making pictures
like Fig. \ref{fig:0}.

During most of the work we combined the data from two sources on
$t=0$ and $t=16$. This stabilizes the extraction of the masses 
significantly for $n=0$. However, as we include more eigenmodes
in the sum over source positions, this gain decreases and becomes
almost negligible for $n=20$. At light quark masses an ``all to all'' calculation 
restricted to low eigenmodes is competitive in cost of reducing fit uncertainty
with the use of full propagators from several source points per lattice.

\end{document}